\documentclass[twoside]{article}
\usepackage{latexsym}
\usepackage{amsmath}
\usepackage{amsfonts}
\usepackage{amssymb}
\usepackage{times}
\usepackage{graphicx}
\newcommand{\be}{\begin{eqnarray}}
\newcommand{\ee}{\end{eqnarray}}
\newcommand{\nn}{\nonumber \\}

\newcommand{\mytil}{\tilde{~}}

\newcommand{\openone}{\mbox{1\kern -0.25em I}}
\newcommand{\openK}{\mathbb K}
\newcommand{\openZ}{\mathbb Z}
\newcommand{\openR}{\mathbb R}

\newcommand{\openC}{\mathbb C}
\newcommand{\Id}{\mbox{I\kern -0.1em d}}
\newcommand{\con}{\mbox{$\,$\rule{1ex}{0.4pt}\rule{0.4pt}{1ex}$\,$}}
\newcommand{\dwedge}{{\,\dot{\wedge}\,}}
\begin{document}   
\title{Clifford geometric parameterization\\
of inequivalent vacua}
\author{Bertfried Fauser\\
Universit\"{a}t Konstanz\\
Fakult\"{a}t f\"{u}r Physik\\
Fach M 678\\
78457 Konstanz {\bf Germany}\\
Electronic mail: Bertfried.Fauser@uni-konstanz.de
}
\date{Feb 14, 2001}                                                  
\maketitle

\begin{abstract}
We propose a geometric method to parameterize inequivalent vacua
by dynamical data.
Introducing quantum Clifford algebras with arbitrary bilinear forms we
distinguish isomorphic algebras --as Clifford algebras-- by
different filtrations resp. induced gradings. The idea of a
vacuum is introduced as the unique algebraic projection on the
base field embedded in the Clifford algebra, which is however
equivalent to the term vacuum in axiomatic quantum field theory
and the GNS construction in 
$C^*$-algebras. This approach is shown to be equivalent to
the usual picture which fixes one product but employs a variety
of GNS states. The most striking novelty of the geometric
approach is the fact that dynamical data fix uniquely the vacuum and that
positivity is not required. The usual concept of a statistical quantum 
state can be generalized to geometric meaningful but non-statistical, 
non-definite, situations. Furthermore, an algebraization of states takes
place. An application to physics is provided by an $U(2)$-symmetry producing 
a gap-equation which governs a phase transition. The parameterization 
of all vacua is explicitly calculated from propagator matrix elements. 
A discussion of the relation to BCS theory and Bogoliubov-Valatin 
transformations is given.
\end{abstract}

\section{\protect\label{SEC-1}Introduction}

\subsection{Prologue}

The core material of this paper, mainly done in 1997, originated from an 
analysis of the $C^*$-algebraic considerations of Kerschner \cite{Ker1,Ker2}
and had a major impact on the progress of a whole set of further work,
supporting the need of an exposition of the method itself.  
Kerschner reflected the need of a more involved concept of (ground) state
or vacuum in quantum field theory. Due to Haag's theorem an interacting 
quantum field theory (QFT) cannot be treated in Fock space. Since 
Kerschner and the group of Prof. Stumpf were involved in composite 
calculations, see \cite{StuBo}, this problem was serious. However, since
regularization issues made it necessary to perform calculations in an 
indefinite state space $C^*$-algebraic methods are not appropriate because
they are tied to positive definiteness which causes problems. It proved to
be a very fruitful concept to turn this considerations into a Clifford
algebraic picture. A reformulation of QFT was done in \cite{Fau-diss}. It 
was necessary to enlarge the concept of Clifford algebra to arbitrary 
bilinear forms which are not necessary symmetric, see e.g. 
\cite{Fau-man,AblLou,Ozi-Grass-Cliff}. Later on, Dirac theory \cite{Fau-ach},
composite calculations in quantum electro dynamics \cite{Fau-pos} 
and normal-ordering in quantum field theory \cite{Fau-ver,Fau-wick} have been 
treated. Surprisingly several singularities which usually occur during 
re-ordering in QFT vanished due to the correct algebraic treatment. A general 
method to deal with QFT in Clifford algebraic terms was given in \cite{chap4}.
The method allowed to generalize some concepts of QFT. Especially it was
shown \cite{Fau-qspin,Hecke1,AblFau,Fau-manin} that $q$-symmetry can be
treated within this formalism. A plain approach to the notions and
peculiarities of quantum Clifford algebras was given in
\cite{FauAbl}. Some further achievements will be discussed in the
epilogue. All this applications have its foundation in the method which
is presented in the present work. A low dimensional model which shows all
peculiarities of QFT while being still fully accessible to calculations
seemed to be the perfect playground to demonstrate the method since a
comparison to $C^*$-algebraic methods is also easily possible.

We want to add a few remarks on notational and conceptual issues about the 
concept of {\bf vacuum} used here:
\begin{itemize}
\item
The term vacuum could be called effective or generalized vacuum to 
express some differences with conventional settings. But from a 
physicists point of view it is simply the unique state which is used
to generate a state space and calculate expectation values. This is the 
physical vacuum of the theory.
\item
If the vacuum is seem as a ground state or a cyclic lowest weight vector
one assumes a Hamiltonian and some annihilation operators which annihilate 
this state. Since we do not choose a particular Hamiltonian but deal with a
whole class of dynamics invariant under a group action, e.g. $U(2)$, one
cannot characterize vacua in this way. Moreover, in non-Fock situations
one cannot expect to find annihilators in terms of the field operators, 
see below.
\item
A basic difference between the $C^*$-algebraic approach an our method
is that in $C^*$-theory one deals with a state {\it space\/}, i.e. a 
convex set of positive linear functionals. An operator equation can in 
principle be evaluated with any state from this set. But in our approach 
the method yields an {\it unique\/} state and no freedom is possible.
This is a major achievement since a unique representation is chosen from 
a possibly infinite set. The theory is thus more predictive. A unique state 
fixes properties of a quantum field theory as e.g. symmetry breaking 
or forming of a condensate. This connection is established via the dynamics 
of the theory which allows to speak about a dynamically chosen representation. 
\item
The last point has to be supplemented by the observation that our vacuum
state cannot be addressed as a thermalized state as it is done in thermo 
field dynamics.
\end{itemize}

\subsection{Motivation}

Quantum physics arose from the necessity to generalize
coordinates and coordinate functions to operators and operator
valued functions. The historical development of quantum
mechanics led to the concept of a Hilbert space, a linear space
with a positive definite inner product, or scalar product, usually
over the complex number field \cite{vonNeu,Dirac}. The so-called
observables are then no longer differentiable functions on a
phase-space, but bounded, hermitean operators on an appropriate Hilbert
space \cite{vonNeu}. We want to emphasize here, that operators are not
observables in their own right. From an empirical point of view,
lets assume a measurement yields a real number. A measurable
quantity is thus a basis dependent expectation value e.g. a
matrix element {\it or}\/ an invariant object which is not basis 
dependent, e.g. an eigenvalue. Only in the positive definite case
of a Hilbert space, are we able to reconstruct the operators up to
isomorphisms, hence their invariants, uniquely from such data;
finite dimension or separability is also assumed. This {\it a
posteriori}\/ identification allows us to address the bounded
operators on a Hilbert space as observables. We will be
concerned here with a geometric relation between operators and their
invariants. 

In quantum field theory, however, one fails to find such a
direct connection. The canonical (anti) commutation
relations CCR (CAR) of a continuum of variables --the
space-time continuum has to be considered as an index-- do
posses infinitely many representations. One does therefore lose
the information about a particular representation when passing
from expectation values to operators and possibly an operator
dynamics. On the other hand, the freedom in choosing a representation
can be used to treat fields at finite temperature or within different
phases and provides therefore an ideal tool in modelling physical
systems. Furthermore, it is known from Haag's theorem, that
interacting QFT's cannot be described correctly within Fock
space \cite{Haag,Araki}. It is thus of utmost importance to have
a {\it constructive}\/ way to handle non-Fock situations. Troubled
with the requirement of positivity, in the relativistic setting,
these situations are usually handled with complicated inductive
limits in a $C^*$-algebraic approach \cite{Haag-buch}. Our
geometric approach will allow such intriguing situations in the
finite dimensional case also, which might be used to circumvent
convergence problems. 

The need of positivity and definiteness results from another
point also. Quantum mechanics was developed within a background
of positivism and it was welcomed by its founders to introduce a
statistical and {\it a priori}\/ unquestionable element into the
theory --the chance. Especially Bohr and the so-called
Copenhagen interpretation developed by Born took this route.
Nevertheless, already in the thirties and mostly connected with
the appearance of the electron spin, geometrical concepts were
investigated \cite{Fock}. We will return to geometric
conceptions which include, however, the QM case as a subset.

In this paper we give a geometric interpretation of vacuum
states and their relation to symmetry and the dynamics under
consideration within a Clifford algebraic framework. However, 
much effort is spent to show that in the positive case of QM
our approach reproduces the common results, even if it is in
some sense more general. This method will include a geometric
interpretation of ``quantization'' also. We will concentrate
on finite dimensional examples i.e. fermions for reasons of
clarity and to avoid convergence problems, which do not belong
to the geometric picture. However, the method can be directly
applied to quantum field theory 
\cite{Fau-pos,Fau-ver,Fau-diss}. Furthermore, one should note, 
that in our framework all the interesting phenomena as phase 
transitions etc. can be handled with finitely many particles. 

The striking advantage of the geometric picture is that it is
neither connected to positivity nor to definiteness. If we
artificially fix the situation in the positive definite case, which
can be handled in the usual way, we are ready to change to the 
geometric picture and afterwards extend the theory to formerly
meaningless situations. Since in quantum field theory ghost
states --of negative and zero norm-- are presently used e.g. in
QCD, we expect a clarification of these situations by our
approach. 

An algebraic definition of the term vacuum is needed, which 
includes conventional vacua, but is accessible also to the general
algebraic picture. The most important feature is invariance
under a symmetry group, which is in axiomatic QFT the Lorentz
group. We will use for simplicity the nonrelativistic case and employ 
the $U(2)$. The construction of states will be done by an analogy 
to $C^*$-algebraic GNS states \cite{Murphy}, which however are 
generalized to indefinite cases by Clifford geometric methods.
As a major outcome, which distinguishes our method from any other 
current approach, the Clifford algebra will provide an unique 
vacuum for every dynamical situation, enabling us to connect 
propagator data to vacuum states.

Since we are presenting a geometric approach to the concept
of states and furthermore their algebraic treatment, we have to
compare conventional and new calculations. Therefore, in section
\ref{SEC-2} an explicit treatment of the one dimensional extended 
Fermi oscillator is given, first to fix the notations and second
for didactical reasons, since this is the model where old and new
approaches share most of their features.

Section \ref{SEC-3} is devoted to the development of the
geometric theory in the finite dimensional case, which however
can easily be used in QFT by analogy. A resume of quantum
Clifford algebras with arbitrary bilinear forms is given in
section \ref{SEC-3.1}. Sections \ref{SEC-3.2} and \ref{SEC-3.3}
treat once more the Fermi oscillator, simply by translating the
results from \ref{SEC-2} and only thereafter turning to a
radical geometric point of view. It seemed to be necessary to
perform this transition in two steps, to be as clear as
possible. A redundancy occurs only in the results, but not in
the novel points of view. Section \ref{SEC-3.4} provides a
treatment of the Fermi oscillator with two degrees of freedom,
first giving the results of ECAR calculations and then a complete
discussion within the geometric picture. The four main results 
given there are a description of inequivalent vacua in a 
parametric and constructive form, the explicit calculation of 
the functional dependency of these parameters from the matrix 
elements of the propagator, a `gap-equation' which distinguishes 
different phases, and a complete classification of all vacuum states 
by geometrical data. 

The \ref{SEC-4}th section gives the connection of our approach to 
BCS theory and Bogoliubov-Valatin transformations as a relation to 
results from literature. Furthermore, it speculates, how this approach 
may be related to the Higgs mechanism and ghost fields. 

\section{\protect\label{SEC-2} The extended Fermi
oscillator}

Let us consider the ordinary Fermi oscillator with one
degree of freedom. This very trivial system serves us finding
the relevant relations by explicit calculations.

The CAR algebra of one degree of freedom is generated by
${\openone}$, and $a, a^\dagger$, where the relations
\be
&aa^\dagger + a^\dagger a= \openone & \nn
&aa = 0 = a^\dagger a^\dagger&
\ee
hold. We will deal with the real field for simplicity, but a
complexification could easily be performed by $CAR_{\openC} = \openC
\otimes CAR_{\openR}$, where bilinear forms have to be changed
into sesquilinear ones.

We find an algebraic basis of the algebra which spans
a (real) linear space $FCAR$ of dimension $2^n = 4$ -- using the 
forgetful functor $F$. Generally we will write $FA$ for the linear
space underlying the algebra $A$ --  
\be
&\{e_i\}_{i=1}^4 = \{ \openone, aa^\dagger, a , a^\dagger \}.&
\ee 
A {\it state}\/ $\omega$ on this algebra is an element of the 
linear forms $lin[FCAR,\openR] \simeq FCAR^*$, where $*$ does 
indicate the dual with respect to a canonical pairing. It is
usually supposed to be linear, positive and normed. Since $FCAR$ 
and $FCAR^* $ are isomorphic as linear spaces, any dual element
can be parameterized by its action on the basis elements in $FCAR$. 
Introducing an {\it independent}\/ ``particle number'' operator or
even better a grading by 
\be\label{Q-act}
   [Q,a] &=& -a\nn
{} [Q,a^\dagger] &=& +a^\dagger \nn
   Q &\not\in& {\bf Alg}(a,a^\dagger),
\ee
we define the Extended CAR (ECAR) algebra \cite{Majid}. Regarding 
$Q$ as an independent quantity, allows us to have circumvented 
von Neumann's uniqueness theorem and provides inequivalent vacua in 
the finite dimensional case also. This is also the place where 
deformations, especially the $q$-business, enters the setting
\cite{Hecke1,Fau-qspin}.

A CAR algebra subjected to the action of a group $G$ here $SO(2)$
resp. $U(1)$ generated by a set of generators e.g. $Q$, is called 
a CAR-dynamical system. 

A {\it vacuum state $\omega$}\/ is defined to be a state invariant
under the action of the group G here a $U(1)$ introduced by $Q$. The 
invariance condition reads
\be\label{inv-cond}
& \omega (A)=\omega (QA) \quad (=\,\, \omega (AQ)\,). &
\ee
This is exactly the definition of a vacuum due to axiomatic
quantum field theory \cite{SW}. An emphasis is laid on symmetry
principles and not on step operators --creation and annihilation
operators-- or ground states which are annihilated by the later.
See also the discussion of Bogoliubov-Valatin transformations
and of the BCS vacuum below.

In the case of axiomatic QFT \cite{Wightmann}, the desired group
would be the Lorentz group, however, we deal with $U(1)$ and $U(2)$ 
for simplicity. The $U(2)$ example is already of physical interest
and models superconductivity. Furthermore, if a 
particle interpretation is desired, the vacuum state
has to be a highest (lowest) weight state of a specific representation
of the group acting in our system. This is in fact the GNS state 
$\mid \Omega_k> \cong \mid 0 >_{GNS}$ of a {\it specific}\/ 
representation $\pi$. However, we also deal with indefinite situations.
A lowest weight state is called {\it ground state}\/ for physical
systems and will be denoted as $\mid 0>$. It obeys the 
Fock-like condition $X^- \mid 0> =0$ for some ladder operator 
$X^-$. Furthermore, from (\ref{inv-cond}) we can conclude that a 
vacuum state has no nontrivial quantum number w.r.t. the chosen 
group. State and group action define an ECAR dynamical system, 
if additionally a Hamiltonian element is specified. However we will
see that it is sufficient to specify the propagator of the theory 
{\it as a parameter}\/ which fully specifies the vacuum properties.
This relation leading to multiple Hamiltonians bearing the same 
vacuum structure is not injective. In the sequel, we deal exclusively
with vacuum states denoting them loosely as $S^0(ECAR) = \{ \omega_k \}$,
with $k$ an index set. 

This means that we seek for vacuum states, which are defined to 
posses no nonzero $G$ quantum numbers. Together with the invariance 
condition (\ref{inv-cond}) which leads to $\omega(AQ)=0$, $\forall A 
\in ECAR$, we have to fulfil in our case:
\be\label{ECAR-state}
i)  & \omega(\openone) = 1 \hfill   & \mbox{normalization}\nn
ii) & \omega(a)=0=\omega(a^\dagger) & \mbox{due to the grading}\nn
iii)& \omega(aa^\dagger)=\nu
\ee
where $\nu$ is a real parameter. The positivity requirement
restricts $\nu$ by $\nu > 0$ and $\omega(a^\dagger
a)=\omega(\openone-aa^\dagger) = 1-\nu > 0$ which results in $1
> \nu > 0$. Hence we can parameterize all linear, normed,
positive vacuum states $S^0(ECAR)$ by a single real quantity $\nu
\, \in ]0,1[$. We write $\omega_\nu$ for such {\it vacuum states}\/
 in $S^0(ECAR)$. 

We can now calculate $\nu$-dependent pseudo representation matrices
by using $\omega_\nu$ as linear form and the $\{e_i\}$ basis. With
$A \in ECAR$ we have 
\be
[\mu_{ij}^\nu(A)] &=& [\omega_\nu(e_i^\dagger A e_j)]
\ee
e.g.
\be\label{mu}
[\eta_{ij}^\nu] = [\mu_{ij}^\nu(\openone)] = 
\begin{array}{|cccc|}
1 & \nu & 0 & 0 \\
\nu & \nu & 0 & 0 \\
0 & 0 & 1-\nu & 0 \\
0 & 0 & 0 & \nu 
\end{array}
\hspace{1truecm}
[\mu_{ij}^\nu(a)]=
\begin{array}{|cccc|}
0 & 0 & 0 & \nu \\
0 & 0 & 0 & \nu \\
1-\nu & 0 & 0 & 0 \\
0 & 0 & 0 & 0 
\end{array}
.
\ee
This is not a homomorphism and the algebra product is {\it not}\/ mapped 
onto the matrix product, so $\mu$ is not a representation, which can be
seen from 
\be
\omega_\nu(e_i^\dagger ABe_j) 
&=& \omega_\nu(e_i^\dagger Ae_k)b_{jk} \nn
&=& \omega_\nu(e_i^\dagger Ae_r)\eta^{-1}_{rl}\eta_{lk}b_{jk} \nn
&=& \omega_\nu(e_i^\dagger Ae_r)\eta^{-1}_{rl} b_{jk} 
     \omega_\nu(e_l^\dagger e_k) \nn
&=& \omega_\nu(e_i^\dagger Ae_r)\eta^{-1}_{rl}\omega(e_l^\dagger Be_j) \nn
\ee
or equivalently
\be
\mu(AB) &=& \mu(A)\circ \eta^{-1}\circ \mu(B).
\ee
As long as $\eta$ is nonsingular and positive, which is not the case
in Fock or dual Fock space or for values of $\nu 
\not\in ]0,1[$, see below, one could proceed to a $*$-representation 
$\pi$ by setting
\be
\pi^\nu &:=& \eta^{-1/2} \circ \mu^\nu \circ \eta^{-1/2},
\quad \nu \in \, ]0,1[.
\ee
The two extremal and singular cases $\nu=0,1$ reduce the
situation to usual Fock $F$, $\pi^1(a) \mid 0>_F=0$ and dual Fock 
$F^*$, $\pi^0(a^\dagger) \mid 0>_{F^*} =0$ space after factoring out
a null-space. A connection between them is achieved by renaming 
$a,a^\dagger$ to $a^\dagger,a$. In general, if
$\nu\not=0$ or $1$, the representations are no longer reducible
and remain to be four dimensional. However, a decomposition into Fock 
and dual-Fock spaces can be performed using mixed states. Note once
more, that $Q\not=Q(a,a^\dagger)$ is an independent quantity. However, 
from the group action (\ref{Q-act}) we see that $(ad_Q)^2=Id$ on 
$V = <a,a^\dagger >$ and the basis set is only $\{ e_i\}_{i=1}^4 
\oplus \{ e_i \}_{i=1}^4 Q$.

At the end of this section, we give a further possibility to
describe the $\nu$-dependent vacua, which is needed later. 
From a decomposition of unity into primitive projectors we have
\be
\openone &=& P_F + P_{F^*}
\ee
and the decomposition
\be\label{omega-nu}
\omega_\nu(A^\dagger B)&=&\omega_\nu(A^\dagger 
   [P_F + P_{F^*}] B) \nn
 &=& \omega_\nu(A^\dagger P_F B) + \omega_\nu(A^\dagger P_{F^*} B)\nn
 &=:& \nu \omega_F(A^\dagger B) +(1-\nu)\omega_{F^*}(A^\dagger B),
\ee
where $\omega_F$ and $\omega_{F^*}$ are the extremal Fock and
dual Fock states. If one likes to proceed to a diagonal
decomposition it is convenient furthermore to introduce GNS
ground states $\vert\Omega_\nu>$. This results up to a relative 
phase in
\be
\omega_\nu(A)&=& <\Omega_\nu\mid \pi_\nu(A) \Omega_\nu>\nn
&=&  \nu <\Omega_F\mid \pi_F(A) \Omega_F> 
+ (1-\nu)<\Omega_{F^*}\mid \pi_{F^*}(A) \Omega_{F^*}>\\
\pi_\nu(A) &=& \pi_F(A) \oplus \pi_{F^*}(A)\\
\vert\Omega_\nu> &=& \sqrt{\nu}\vert \Omega_F> \oplus 
              \sqrt{1-\nu}\vert \Omega_{F^*}>.
\ee
Since Fock and dual Fock situations are connected by a simple
relabelling of generators, the distinction seems to be a commonplace. 
But if more generators are involved other representations come into
play, furthermore the concrete expression of $Q$ depends already on this
choice. However, one should be careful in transforming generators,
since mathematical features such as simplicity of the generated
multiplicative groups may depend drastically on such changes, 
see \cite{CM}.

Note, that the GNS states can be built only in the positive setting.

\section{\protect\label{SEC-3} Clifford geometric approach} 

\subsection{\protect\label{SEC-3.1} Nonsymmetric bilinear
forms in Clifford algebras} 

To be able to describe the same structure as in the previous
section, it is necessary to use a precision of the
term Clifford algebra which might be called Clifford algebra
of multivectors according to \cite{Ozi-multivectors} or quantum 
Clifford algebras \cite{FauAbl}. A detailed treatment can be found 
in \cite{chap4,Hecke1,FauAbl}, we give only the
notations and results necessary for our models. Since a
Clifford algebra is --in a functorial sense-- the natural
algebra of a quadratic space ($V$,$Q$), a pair of a linear
space $V$ and a quadratic form $Q$. One can define a functor
$CL$ from the category of quadratic spaces into the category of
unital associative algebras \underline{${\bf alg}$}. But bilinear
forms associated with quadratic forms are necessarily symmetric
if the characteristic of the base field is not equal to 2. The
polar bilinear form of a quadratic form $Q$ is defined with $x,y
\in V$ as
\be
2B_p(x,y)&:=& Q(x+y)-Q(x)-Q(y).
\ee
The Clifford map $\gamma : V \hookrightarrow CL$ results then in
\be
\gamma_x\gamma_y + \gamma_y\gamma_x &\simeq& 
 xy+yx \, =\, 2 B_p(x,y),
\ee
where we identify $x\in V$ and $\gamma_x \in CL$. The Clifford
product of elements from $V$, written without sign, can be
decomposed as \cite{MRiesz}
\be
\gamma_x\gamma_y &=&B_p(x,y) + \gamma_x \wedge \gamma_y,
\ee
where the wedge is an antisymmetric Grassmann product. In
writing this decomposition, we have used an isomorphism of the
linear spaces $FCL$ and $F\bigwedge V$, the linear space of
Grassmann multiforms. This can be used to construct the Clifford
algebra directly from the Grassmann algebra via Chevalley
deformation \cite{Che}. The so defined product on $V$ can be
lifted on the whole algebra $CL$, see (\ref{con}).

Moreover, we are no longer restricted to a symmetric bilinear
form. One can define the left contraction on $V$
\be
\gamma_x \con \gamma_y &=& B(x,y) \, \simeq\, <x\mid y>
\ee
in a completely arbitrary way. $x\con$ is then $\in V^*$ and
$\gamma_x \con$ its image in $CL$, see
\cite{Ozi-Grass-Cliff,Fau-man,Fau-ver,Fau-pos,AblLou,Fau-diss,Fau-ach,Lou}.
Contrary to the associative wedge product, the contraction or
inner product is nonassociative. For multiforms we have to
define with $x_i,y_j \in V$ and dropping from now on the
injection $\gamma : V \hookrightarrow CL$
\be\label{con}
i)   && x_i \con x_j = B(x_i,x_j) \simeq <x_i \mid x_j> \\
ii)  && x_i \con(y_0\wedge\ldots\wedge y_n) = \sum_{r=0}^{n}
     (-1)^r (x_i \con y_r)(y_0\wedge\ldots\wedge y_{r-1}\wedge
            y_{r+1}\wedge\ldots\wedge y_n) \nn
iii) && (x_1\wedge\ldots\wedge x_n)\con (y_1\wedge\ldots\wedge y_m)=
        x_1\con (x_2\con (\ldots (y_1\wedge\ldots\wedge y_m)\ldots))
        \quad n\leq m .\nonumber
\ee
We have however to remark, that the grading inherited from
the Grassmann multivector structure is not unique. One could
also define another contraction, which is symmetric and put the
antisymmetric part into the Grassmann bi-form
\be
xy &=& x \con_B y + x\wedge y = B(x,y) + x\wedge y \nn
   &=& x \con_G y + x\dwedge y = G(x,y) + x\dwedge y.
\ee
Here $B=G+F$ is decomposed into symmetric and antisymmetric
part, we have thus
\be
x \dwedge y &=& F(x,y)+ x\wedge y,
\ee
which clearly exhibits the different gradings. It was shown,
that exactly this change of multivector structure is done when
normal ordering, that is Wick ordering w.r.t. $F$, is performed
in quantum (field) theory \cite{Fau-wick}. Even in nonlinear, resp. 
self-interacting, theories no singular additional terms arise
when this change is treated algebraically correct \cite{Fau-ver}.

In the sequel, we will give the bilinear forms explicitly in
terms of a basis and use either the common wedge multiforms or
monomials in Clifford generators. The Clifford monomials
constitute a further possibility to introduce a grading on $CL$.

\subsection{\protect\label{SEC-3.2} Fermi oscillator in Clifford
terms} 

We start from a two dimensional $(n=2)$ hyperbolic space $H$,
where we introduce the following particular basis
$\{e_i\}=\{e_1,e_2\}$ and symmetric(!) bilinear form
\be
[B(e_i,e_j)] &=& [\frac{1}{2}\delta_{i,n+1-i}] = 
\begin{array}{|cc|} 0 & \frac{1}{2} \\ \frac{1}{2} & 0
\end{array}\, .
\ee
We construct furthermore the Clifford algebra $CL(H,B)$, which
leads to the commutation relations identifying from now on
$\gamma_{e_i} \simeq e_i$ and $\openone \simeq 1$
\be
e_i e_j + e_j e_i &=& B(e_i,e_j) + B(e_j, e_i) + e_i \wedge e_j
+ e_j \wedge e_i \nn
&=& 2B(e_i,e_j).
\ee
If we identify $e_1$ with $a$ and $e_2$ with $a^\dagger$, we
notice that this algebra is isomorphic to the CAR algebra of
the previous section. The relations then read
\be
&aa^\dagger + a^\dagger a = B(e_1,e_2)+B(e_2,e_1) =
\frac{1}{2}+\frac{1}{2} = 1& \nn
& aa=B(e_1,e_1)=0=B(e_2,e_2)=a^\dagger a^\dagger,
\ee
where the second line holds because of $a\wedge a=0 = a^\dagger
\wedge a^\dagger$. To establish the full correspondence, we need
to define the $\dagger$ antiautomorphism. This can be done using
the main antiinvolution $\mytil$ of the Clifford algebra, also
called reversion, defining the transformation $\underline{d}$
with help of the
\be 
p_i := (e_i - e_{n+1-i})
\ee
as
\be
\underline{d}(X)&:=&\prod_{i=1}^{m=n/2=dim V/2}
(- p_i) X\mytil p_i = (-)^m p_1 p_2 \ldots p_m X\mytil p_m
\ldots p_1. 
\ee
It follows $\underline{d}(XY)= \underline{d}(Y)
\underline{d}(X)$. In our case, we have 
\be
\begin{array}{c} \mbox{\underline{$d$}}(e_1)=e_2 \\ 
                 \mbox{\underline{$d$}}(e_2)=e_1
\end{array} 
&\Leftrightarrow&
\begin{array}{c} \mbox{\underline{$d$}}(a)=a^\dagger \\
                 \mbox{\underline{$d$}}(a^\dagger)=a .
\end{array} 
\ee
Contrary to the ECAR case, where we had a correspondence of
$FCAR$ and the dual elements $FCAR^*$, where each of the later
constitutes a linear form, we have a canonical choice for a
linear form in the Clifford algebra. We may define the scalar
part $<.>_0$ of a Clifford number by projecting onto the field
of scalars embedded in $CL$: 
\be
<.>_0 &:& CL \mapsto \openK\nn
<A>_0\,\, = \, <A>_0^\wedge &=& <\alpha\openone +\alpha_{ij}
e_i\wedge e_j + \ldots >_0 \,\, =\, \alpha.
\ee
We drop the wedge superscript when no confusion can occur
which wedge is used to expand the Clifford numbers.
Hence, the scalar part picks out the coefficient in front of the
identity when A is expanded into {\it Clifford multivectors}\/
w.r.t. {\it a certain Grassmann product}.

Now, the states $\omega_\nu$ of the ECAR algebra shall be related
to the single linear form of the Clifford algebra. We therefore
use the representation theory of algebras. An irreducible, faithful
representation can be obtained by a left regular representation
of the simple algebra on a minimal left ideal generated by a
primitive idempotent element. A good choice is either
$P_{F^*}=aa^\dagger  = e_1e_2$, or $P_F=a^\dagger a = e_2 e_1$
because of 
\be
aa^\dagger aa^\dagger&=& -a^2 (a^\dagger)^2+aa^\dagger =
aa^\dagger \nn
a^\dagger aa^\dagger a&=& -(a^\dagger)^2 a^2 + a^\dagger a =
a^\dagger a.
\ee
Furthermore, the ECAR relations show that the two projectors are
a decomposition of unity
\be
\openone &=&P_F+P_{F^*}=aa^\dagger + a^\dagger a.
\ee
We can define now the Fock and dual Fock vacua by letting
\be
\omega_F(X) &:=& <2P_F X>_0 \nn
\omega_{F^*}(X) &:=& <2P_{F^*}X>_0,
\ee
which are identical with the above defined forms, as can be seen
by calculating their values on the generators. Since the factor 2
stems from the ungeometrical normalization of the ECAR relation,
we would prefer the geometrical appropriate relation
\be
\hat{a}_i \hat{a}_j^\dagger + \hat{a}_j^\dagger \hat{a}_i &=& 2\delta_{ij}
\openone = 2B_p(\hat{e}_i,\hat{e}_{n+1-j}),
\ee
remembering that $a^\dagger_i=e_{n+1-i}$. The $\nu$-dependent vacua
are then given as in (\ref{omega-nu}) by  
\be
\omega_\nu(X) &:=& v \omega_F(X) +(1-\nu) \omega_{F^*}(X) \nn
&=& <[\nu 2P_F+(1-\nu)2P_{F^*}] X>_0.
\ee
The grading operator this time, formulated within the algebra, is
necessarily $\nu$-dependent, because of the requirement
$\omega_\nu(Q)=0$ and found to be
\be
Q &:=& (\nu-\frac{1}{2})-e_1\wedge e_2 = \nu\openone-aa^\dagger.
\ee
All results of the ECAR theory can be obtained now in purely
Clifford algebraic terms, e.g. the $\mu$-matrices as
$[\mu_{ij}^\nu(A)] \simeq [<[\nu P_F+(1-\nu)P_{F^*}]
\mbox{\underline{$d$}}(E_i) A E_j >_0]$ where the monomials $E_i$
run in $\{\openone, e_1 e_2, e_1, e_2 \}$. However, since we do not
need a representation, but used the basis only to show the equivalence
to the above discussed C$^*$-algebraic description, we need not bother 
about further transformations and positivity requirements.

Since we have up to now hardly done more than to reformulate the
ECAR results, we have to face the question why we prefer a
Clifford geometric approach. Even the correspondence to a single
linear form, sic a trace, is common to physicists, who are used
to working with matrix representations. Every state on a finite
dimensional real or complex (E)CAR algebra can be written as
\be
\omega_\rho(X) &=& \frac{1}{dim V} Tr(\rho \pi(X)),
\ee
where $V$ is the representation space, $\pi : (E)CAR \mapsto
End(V)$ a representation and $\rho$ the density matrix. 
However, a representation is implicitly needed in such a 
construction. In the next subsection we utilize the Clifford algebra
in a new and geometric way. This step, taking the Clifford geometric
character fully into account, is the doorway to generalize the
situation thereafter {\it beyond}\/ ECAR possibilities.

\subsection{\protect\label{SEC-3.3}Clifford geometric Fermi
oscillator}  

In the previous subsection we have modelled the same situation as
in the ECAR algebra by simply translating them into Clifford
terms. It is however clear, that the full power of a
mathematical tool can only be achieved if its generic abilities
are used. Hence, we will give a further approach to the
Fermi oscillator which relies fully on Clifford geometric
methods. 

Let $B_\nu$ be a bilinear form on the space $V$, ($dim V=2$)
which generates $CL(V,B_\nu)$. In a distinguished basis
$\{e_i\}=\{e_1,e_2\}$ we have
\be
[B_\nu] &=&
   \begin{array}{|cc|} 0 & \nu \\  1-\nu & 0 \end{array}
=  \begin{array}{|cc|} 0 & \frac{1}{2} \\  \frac{1}{2} & 0 \end{array}
  +\begin{array}{|cc|} 0 & -\frac{1}{2}+\nu \\ 
                      \frac{1}{2}-\nu & 0 \end{array} \nn
&=& [G]+[F_{-1/2+\nu}] = [G]+[F_{\nu^\prime}].
\ee
The normalization is chosen in such a way, that $\nu$ will have
the same values as above for positive solutions, $\nu \in\,
]0,1[$. Defining a contraction $\con_\nu$ on $V$ and lifting it
to multivector arguments as in (\ref{con}), we come up with the
Clifford algebra $CL(V,B_\nu)$. Contraction, Clifford and wedge
products are related by 
\be\label{d-wedge}
x \con_\nu y &\equiv& x \con_{B_\nu} y = B_\nu(x,y) \nn
xy &=& x \con_\nu y + x \wedge y \nn
   &=& x \con_G y + x \con_{F_{-1/2+\nu}} y + x \wedge y \nn
   &=& x \con_G y + x \dwedge y.
\ee
Obviously, $\wedge$ and $\dwedge$ define different multivector
structures in $CL$. Indeed, we had already mentioned that $CL$
is only $\openZ_2$-graded and bears {\it no natural}\/
multivector structure. In other words, $CL$ does not depend on
the $\openZ_n$-grading and we have a Clifford algebra
isomorphism, that is the Wick theorem 
\cite{Fau-ver,Fau-diss,FauAbl,Fau-wick},
$CL(V,B_\nu) \simeq CL(V,G)$, where $G$ is symmetric
and does correspond directly to a quadratic form or say a 
``quantization''. This can be seen by inspection of the commutator 
relations which turn out to be independent of $\nu$ ($n=dim V$)
\be
e_i e_j + e_j e_i &=& B_\nu(e_i,e_j) + B_\nu(e_j,e_i) \nn
&=& 2 G(e_i,e_j) = \delta_{i,n+1-j} \nn
&=& Q(e_i+e_j) - Q(e_i) -Q(e_j).
\ee
This yields immediately the identification $a=e_1$, $a^\dagger =
e_2$. If we would like to insist on symmetry, we should now
shift $\nu$ to $\nu^\prime=-1/2+\nu$. But to be able to compare
the results with previous sections, we remain with $\nu$. 

Since we noted, that the scalar-part projection depends on the
multivector structure, which is now parameterized by $\nu$, we
have an intrinsic way opened to describe vacuum states. However,
this is now not a question of a dual-isomorphism connecting $V$
and $V^*$, positivity and a statistical interpretation, but
simply a matter of the geometry emerging from an additional
antisymmetric part in the contraction. Moreover, we showed in
\cite{Fau-pos,Fau-diss} that $F$ equals the propagator of the
theory. We have thus found a constructive way to relate the
dynamics i.e. propagator data to representations. 

To demonstrate the equivalence of $<\ldots>_0^\wedge$
--scalar-part of $CL(V,B_\nu)$ without any Fock-projectors
involved-- and $\omega_\nu$ we calculate the values of {\it the 
single}\/ state $S^0(CL(V,B_\nu))$ on the generators (the linear 
form taken element-wise)
\be
<\{\openone, aa^\dagger, a, a^\dagger \}>_0^\wedge
 &=& \{1,\nu,0,0\} ,
\ee
compare with eqn. (\ref{ECAR-state}). Since the Clifford product 
and the anticommutation relations are $\nu$-independent, we have
established an algebra isomorphism. Since we had
implemented the $\dagger$ via a Clifford reversion followed by
a linear transformation $\underline{d}$, we have in fact 
constructed a $*$-isomorphism. One should remark however that
the Clifford reversion does {\it not}\/ respect in general the
multivector structure. This can be seen from 
\be
(e_1 \wedge e_2)\mytil &=& [ e_1 e_2 - B_\nu(e_1,e_2)]\mytil \nn
&=& e_2 e_1 -B_\nu(e_1,e_2) = e_2\wedge e_1 +
B_\nu(e_2,e_1)-B_\nu(e_1,e_2) \nn
&=& -e_1\wedge e_2 -2F_{1/2-\nu}(e_1,e_2).
\ee
On the other hand, the dotted wedge $\dwedge$ defined in (\ref{d-wedge}) 
is stable under reversion \cite{Fau-ach}. This motivated the normal 
ordering procedure in QM and QFT \cite{Fau-ver,Fau-diss}. However, one has 
{\it carefully to distinguish}\/ then the scalarpart projections 
$<\ldots >_0^\wedge$ and $<\ldots >_0^\dwedge$, which result in 
entirely different states. The later constitutes the maximally 
mixed state w.r.t. $\wedge$
\be
<\{\openone, aa^\dagger, a, a^\dagger \}>_0^\dwedge &=&
\{1,\frac{1}{2},0,0\} \quad \Rightarrow\quad \nu=\frac{1}{2}, \quad
(\nu^\prime=0),
\ee
which of course also equals the usual {\it normal ordered}\/ Fock
state w.r.t. $\dwedge$. The state is positive if $\nu\in ]0,1[$, 
which is equivalently written as
\be
-\mbox{det}\, B_\nu > 0.
\ee
The relation between geometry and algebraic norms of such types
were examined in \cite{Fau-man}. In the cases $\nu=0$ and
$\nu=1$, the base space has only a degenerated form, which does
not allow the construction of a universal Clifford algebra.
Finally, we could furthermore calculate once more, to show the 
equivalence to our previous results, the matrices
$[\mu_{ij}^\nu (X)]=[<\underline{d}(E_i) X E_j >_0^\wedge ]$.
One should note, that within the Clifford algebraic approach, 
a representation does not become necessary, since all calculations can 
be performed within {\it abstract}\/ algebra. Hence, we are not troubled 
with the positivity requirement to obtain representations in Hilbert
spaces via the GNS construction. 

\subsection{\protect\label{SEC-3.4} Two degrees of freedom}

As we have already said, the one dimensional system served as a
learning field. In moving to the next dimension, we
will however see some probably unexpected details. This system
is furthermore able to describe physical situations, especially
the occurrence of spin-zero and spin-one can be used for the
purpose of bosonization. 

The two-dimensional system also exhibits differences between the
ECAR and Clifford approach, which goes beyond the simple
restriction of some parameters to gain positivity. The
coherence that {\it all}\/ entities in the theory are formulated
within {\it one}\/ mathematical system, in the multivector
Clifford geometric algebra, unfolds dependencies and relations
which were in principle not visible in the former ECAR setting.
This is a further example of the usability of geometric quantum 
Clifford algebras as claimed by Hestenes \cite{Hes-univ}.

\subsubsection{ECAR results}

Since we fear having already overdone the obvious in the
treatment of the one-dimensional system, we will give in this
section only the definitions and results. Most of them can be
found in \cite{Ker1,Ker2}. 

We define the $U(2)$-ECAR algebra generated by $\{a_\alpha,
a_\beta^\dagger, \openone, S_k, Q \}$, $\alpha,\beta \in \{1,2\}$, 
$k \in \{1,2,3\}$ via
the relations
\be
&\{a_\alpha, a_\beta \} = 0 = \{a_\alpha^\dagger,
a_\beta^\dagger \} \nn
&\{a_\alpha, a_\beta^\dagger \}  = \delta_{\alpha,\beta} \openone.&
\ee
Furthermore, we demand an action of the group $U(2)$, if the
complex ground field is used, by the requirements
\be
& [Q,S_k] = 0, \quad [S_k,S_l] = i\epsilon_{klm}S_m &\nn
& S^\dagger = S,\quad Q^\dagger = Q & \nn
& [S_k, a_\alpha] = \sigma_k^{\alpha \beta} a_\beta , \quad
 [S_k, a_\alpha^\dagger] = \hat{\sigma}_k^{\alpha \beta}
a_\beta^\dagger & \nn
& [Q,a_\alpha] = + a_\alpha,\quad
  [Q,a_\alpha^\dagger] = - a_\alpha^\dagger, &
\ee
where $\sigma_k^{\alpha \beta}$ are the Pauli-matrices and
$\hat{\sigma}_k = -\bar{\sigma}_k = (-)^k \sigma_k$ (the
relations are not independent). Since we do not fix any
dependency on the $S_k,Q$ as functions of the
$\{a_\alpha, a_\beta^\dagger, \openone\}$, we have defined an
Extended CAR algebra, \cite{Zachos}.

A maximal set of pair-wise commuting group elements is given by
$Q,S_3$ and $\vec{S}^2=\sum_{k=1}^3 S_k^2$. We feel very
uncomfortable with this notation, because $S_k$ will be a
multivector aggregate within the Clifford algebra and not a
`vector', but we use it because of readability and since it is a
standard. 

Additionally, we define a distinguished basis $\{ g_i\}$ in terms
of polynomials in the Clifford generators, which are chosen to be
eigenvectors of the commuting operators $S_3,\vec{S}^2,Q$, with
eigenvalues $s_3,s(s+1),q$. This is summarized in table 1, see
\cite{Ker1}. 

\begin{table}[t]
{\hfill
\begin{tabular}[t]{|c|c||c|c|c|}
\hline\hline
\multicolumn{5}{c}{\bf Table 1.} \\
\hline\hline
No: & $A \in ECAR$ & $s_3$ & $s(s+1)$ & $q$ \\
\hline\hline
$g_1$   & $\openone$ & 0 & 0 & 0 \\
$g_2$   & $\frac{1}{2}(a_1a_1^\dagger+a_2 a_2^\dagger)$ & 0 & 0 & 0 \\
$g_3$   & $a_1 a_2 a_2^\dagger a_1^\dagger$ & 0 & 0 & 0 \\
\hline
$g_4$   & $a_1 a_2^\dagger$ & 1 & 2 & 0 \\
$g_5$   & $\frac{1}{2}(a_1a_1^\dagger-a_2 a_2^\dagger)$ & 0 & 2 & 0 \\
$g_6$   & $a_2 a_1^\dagger$ & -1 & 2 & 0 \\
\hline
$g_7$   & $a_1$ & $\frac{1}{2}$ & $\frac{3}{4}$ & 1 \\
$g_8$   & $a_1 a_2 a_2^\dagger$ & $\frac{1}{2}$ & $\frac{3}{4}$ & 1 \\
$g_9$   & $a_2$ & $-\frac{1}{2}$ & $\frac{3}{4}$ & 1 \\
$g_{10}$& $a_2 a_1 a_1^\dagger$ & $-\frac{1}{2}$ & $\frac{3}{4}$ & 1 \\
\hline
$g_{11}$& $a_2^\dagger$ & $\frac{1}{2}$ & $\frac{3}{4}$ & -1 \\
$g_{12}$& $a_1 a_1^\dagger a_2^\dagger$ & $\frac{1}{2}$ & $\frac{3}{4}$ & -1 \\
$g_{13}$& $a_1^\dagger$ & $-\frac{1}{2}$ & $\frac{3}{4}$ & -1 \\
$g_{14}$& $a_2 a_2^\dagger a_1^\dagger$ & $-\frac{1}{2}$ & $\frac{3}{4}$ & -1 \\
\hline
$g_{15}$& $a_1 a_2$ & 0 & 0 & 2 \\
$g_{16}$& $a_1^\dagger a_2^\dagger$ & 0 & 0 & -2 \\
\hline
\multicolumn{5}{p{6cm}}{\medskip

Table 1.: Eigenvectors used as basis states 
and their $U(2)$ quantum numbers}
\end{tabular}\hfill}
\end{table}

The linear space spanned by $\{ g_1$, $g_2$, $g_3 \}$ contains
the candidates for vacuum states. To be precise, $\{g_1,g_2,g_3\}$
are in the kernel of the map $\omega \, :\, F\!A \mapsto \openK$.
One may note the occurrence of a spin-one triplet ($s(s+1)=2$) 
which of course has zero expectation values in Fock-space, of two 
spin one-half ``quartets'' ($s(s+1)=3/4$) of opposite charge or 
``particle number'' and two spin zero eigenvectors of ``particle 
number'' $q=\pm2$.  

In \cite{Ker1,Ker2} the vacuum states were deduced from special
conditions (given below, (\ref{e5})) on this basis. However, one
might notice that these states are defined only up to an
additive, possibly complex constant e.g. $g_5 \simeq
g_5^\prime=g_5 + c\openone$, since the constant does not
contribute to the commutator e.g. $[Q,g_5] \simeq [Q,g_5^\prime]
= [Q,g_5+c\openone ]$. This fact spoils the claim in
\cite{Ker1,Ker2} that one can conclude from $<\Omega\mid g_5
\Omega> = <\Omega\mid \frac{1}{2}(a_1 a_1^\dagger -a_2
a_2^\dagger) \Omega>=0$ the validity of  $<\Omega\mid a_1
a_1^\dagger\Omega> = - <\Omega \mid a_2 a_2^\dagger \Omega> (=
\nu)$. We will nevertheless stay with this restriction to be
able to compare ECAR and Clifford geometric results and impose
the {\it normalizations}\/ of the $g_i$ given in table 1 as
further constraints. However, we will see that the pair
($\nu,w$) is sufficient to parameterize all vacuum states, but
the ECAR and thereby the $C^*$-algebraic approach is then no
longer able to relate them to other data, as to the propagator.
This is a major drawback not apparent in the multivector Clifford
algebraic framework. 

With the basis of table 1 (including normalization), we conclude
that vacuum states can be parameterized by two real variables
$\nu,w$ since we have
\be\label{e5}
\hspace{-0.5cm}
&\omega_{\nu w}(g_1)=\omega_{\nu w}(\openone) = 1& 
\mbox{normalization of $\omega_{\nu w}$} \nn
\hspace{-0.5cm}
&\omega_{\nu w}(g_5) \,\Rightarrow\, \quad \omega_{\nu w}(a_1
a_1^\dagger) = \omega_{\nu w}(a_2 a_2^\dagger) = \nu & \mbox{see
remark above} \nn 
\hspace{-0.5cm}
&\omega_{\nu w}(g_3)=\omega_{\nu w}(a_1 a_2
a_2^\dagger a_1^\dagger) = w & \nn
\hspace{-0.5cm} 
&\omega_{\nu w}(g_4)=\ldots = \omega_{\nu w}(g_{16}) =0 &
\mbox{bec. of nontrivial eigenvalues.}
\ee
If we further require that $\omega_{\nu w}$ is a positive
state, we have to restrict the parameters to
\be
& 1 > \nu > w > 0 & \nn
& w > 2\nu -1, &
\ee
which describes the interior of a simplex, see Fig. 1. It was 
demonstrated in \cite{Ker1,Ker2} that  $\omega_{\nu w}$ can 
be decomposed into three extremal states, once more the Fock 
and dual Fock states with $\omega_F=\omega_{11}$, 
$\omega_{F^*}=\omega_{00}$ and a further uncommon state 
$\omega_E=\omega_{1/2\, 0}$. This yields the decomposition 
\be
\omega_{\nu w} &=& w \omega_{F} +(1-2\nu+w)\omega_{F^*}
+2(\nu-w)\omega_{E}.
\ee
Quasi free states, with vanishing higher correlations \cite{32}
turn out to be described by $\omega_{\nu\, \nu^2}$, a parabola
in the $\nu$-$w$-plane. It is interesting that Bogoliubov
transformations which mix Fock and dual Fock states do
therefore create correlations, but with no contribution of
$\omega_{E}$, see also discussion below.

We end now the discussion, further information on the dynamics
and the thermodynamical behaviour generated by this type of
system can be found in \cite{Ker1,Ker2}. However, one should 
keep in mind that the ECAR treatment is not able to relate the 
parameters $\nu,w$ to dynamical data. 

\subsubsection{Clifford geometric results}

If we use any possible freedom in defining a contraction, the
Clifford algebra provides a richer structure than the ECAR 
algebra. Since we want to stay with a correspondence $\{a_1,
a_2, a_2^\dagger, a_1^\dagger \} = \{e_1, e_2, e_3, e_4 \}$, we
require that the symmetric part of the contraction of the
$\{a_i\}$ and $\{a_j^\dagger\}$, leads to 
\be
\{ a_\alpha, a_\beta \} &=0=& \{a_\alpha^\dagger,
a_\beta^\dagger \} \nn
\{a_\alpha , a_\beta^\dagger \} &=& \delta_{\alpha \beta}
\openone. 
\ee
Canonical quantization turns out to be nothing more than the
specification of an appropriate quadratic form in use
\cite{Fau-pos,Fau-diss}. The antisymmetric part of the
contraction is {\it a priori}\/ arbitrary; we set therefore 
\be
[a_\alpha \con_B a_\beta^\dagger] &=& [e_i \con_B e_{j+2}] =
\begin{array}{|cc|}
q & r \\ s & t \end{array} \nn
{}[a_\alpha^\dagger \con_B a_\beta] &=& [e_{i+2} \con_B e_j ] =
\begin{array}{|cc|}
-q & 1-s \\ 1-r & -t \end{array}\, ,
\ee
and note that from $\{a_\alpha, a_\beta \}= 0 = \{a_\alpha^\dagger,
a_\beta^\dagger \}$ no additional information is obtained from the
antisymmetric part, which finally yields in terms of Clifford generators 
\be
[B(e_i,e_j)] &=& \begin{array}{|cccc|}
0 & u & q & r \\
-u & 0 & s & t \\
-q & 1-s & 0 & m \\
1-r & -t & -m & 0
\end{array}\, .
\ee
A canonical grading operator may be defined by the identity
operator on the space spanned by the $\{a_\alpha \}$ alone
\be
Q^\prime &:=& a_1 a_1^\dagger + a_2 a_2^\dagger.
\ee
The vacuum state is defined as the scalarpart $<\ldots
>_0^\wedge$. We have thus to renormalize $Q^\prime$ by a scalar
additive constant. From
\be
&<Q>_0^\wedge\quad :=\quad <Q^\prime + \alpha\openone
>_0^\wedge=0& 
\ee
we deduce
\be
-\alpha &=& <Q^\prime >_0^\wedge = < e_1 e_4 + e_2 e_3>_0^\wedge \nn
&=&<B(e_1,e_4)+B(e_2,e_3)+e_1 \wedge e_4 + e_2 \wedge e_4
>_0^\wedge \nn
&=&B(e_1,e_4)+B(e_2,e_3)=r+s,
\ee
or
\be
Q&=& a_1 a_1^\dagger + a_2 a_2^\dagger -(r+s)\openone = a_1
\wedge a_1^\dagger + a_2 \wedge a_2^\dagger.
\ee
Observe, that the first expression for $Q$ is independent of the
specific used $\wedge$, but depends on a specific
normalization $-(r+s)$, while the second is independent of the
normalization, but depends on the specific used $\wedge$. From
table 1, we know that $\{g_1,g_2,g_3\}$ might span the vacuum
sector. A vacuum state $\omega_X$ should thereby fulfil
$\omega_X(g_4)= \ldots \omega_X(g_{16})=0$. We calculate $<
\ldots >_0^\wedge$ on all basis elements $\{g_i\}$ to obtain
\be
\{ < g_i >_0^\wedge \} &=& \{ 1,\, \frac{1}{2}(r+s),\,
um-tq+rs,\, q,\, \frac{1}{2}(r-s),\, t,\, \nn
&& 0,0,0,0,0,0,0,0,\, u,\, -m \}\, .
\ee
Because of the choice of normalization in table 1, and the
requirement that $< \ldots >_0^\wedge$ is a vacuum state, we
have to set $< g_4 >_0^\wedge = < g_5>_0^\wedge = < g_6
>_0^\wedge = < g_{15} >_0^\wedge = < g_{16} >_0^\wedge = 0$,
which yields 
\be\label{rel}
& q=0,\quad t=0, \quad r=s,\quad u=0,\quad m=0.&
\ee
The third relation is equivalent to the condition $\omega_{\nu
w}(a_1 a_1^\dagger) = \omega_{\nu w}(a_2 a_2^\dagger) = \nu$ in
(\ref{e5}). But the Clifford treatment uncovers then the
relation of $\nu=r=s$ and $w=um-tq+rs= < g_3 >_0^\wedge$. Since
we have to fulfil (\ref{rel}) consistency is obtained only if
we have $w=rs=\nu^2$ which leads to
\be\label{eq-v1}
\omega_{\nu\, \nu^2 } &=& < \ldots >_0^\wedge, \quad
q=t=u=m=0,\quad r=s=\nu. 
\ee
These are quasi free states and thereby almost trivial. The space
of states is parameterized by one single real number $\nu$, not
a pair ($\nu,w$). To provide a set of positive states, $\nu$ has
to be restricted to $]0,1[$.

The Clifford algebraic examination of the ECAR results unveils
therefore an inconsistency, which breaks up the relation of
contraction data and the parameterization of vacuum states and
quantization. Below, we discuss the connection of the
propagator, contraction and these parameters in Clifford
geometric terms, which is thus {\it beyond}\/ the ability of
ECAR methods. 

Since we want to look for more general situations, we use the
freedom to renormalize the basis $\{g_i\}$, which is defined
only up to constants $g_i^\prime = g_i+c_i\openone$. If we set
\be
g_4^\prime &=& g_4 -q \nn 
g_5^\prime &=& g_5 -\frac{1}{2}(r-s) \nn
g_6^\prime &=& g_6 -t,\nn
g_{15}^\prime &=& g_{15} -u,\nn
g_{16}^\prime &=& g_{16} +m,
\ee
we cancel the unintentional expectation values. We drop the
prime thereafter. The vacuum states can then be parameterized by
two real numbers $\nu,w$, which are of course functions of the
parameters $q,r,s,t;u,m$. One obtains from $g_2$ and $g_3$
\be\label{g23}
< g_2 >_0^\wedge &=& \nu = \nu(q,r,s,t;u,m) = \frac{1}{2}(r+s) \nn
< g_3 >_0^\wedge &=& w = w(q,r,s,t;u,m) =  um - tq + rs.
\ee
If we set $r=-s+2\nu$, we remain with $w=um -tq -s^2+2s\nu$, which
does posses the solutions
\be\label{sol}
& s=\nu \pm \sqrt{\nu^2 -tq +um -w} &
\mbox{\# variables} = 4,\quad 
(q,t,u,m) \, .
\ee
To be able to relate this results with more conventional
approaches, we remark that the propagator, also denoted by $F$,
of the theory was shown to be equivalent to the antisymmetric
part of $B$ in \cite{Fau-pos,Fau-ver,Fau-diss}. This can be seen
as follows  
\be
F_{ij} &=& <\frac{1}{2}[a_i , a_j^\dagger ] >_0^\wedge \quad=\quad
\frac{1}{2} < a_i a_j^\dagger - a_j a_i^\dagger >_0^\wedge \nn
&=& \frac{1}{2} <B(e_i,e_{n+1-j}) - B(e_{n+1-j},e_i)
    + e_i \wedge e_{n+1-j} - e_{n+1-j} \wedge e_i >_0^\wedge \nn
&=& < F(e_i, e_{n+1-j}) + e_i \wedge e_{n+1-j} >_0^\wedge \nn
&=& F(e_i, e_{n+1-j}),
\ee
and hence
\be\label{eq63}
[F_{ij}] &=& [ < \frac{1}{2}[a_i , a_j^\dagger] >_0^\wedge ] =
\begin{array}{|cc|} -r +1/2 & q \\ t & -s+1/2 \end{array}.
\ee
The solutions obtained in (\ref{sol}) yield thus a direct and
{\it constructive}\/ relation between the matrix elements of the
propagator $(q,r,s,t;u,m)$ and the parameterization of the
corresponding vacua $(\nu,w)$.

Since we have to compare our results with ECAR algebraic ones
over the complex number field, we have to take the
($q,r,s,t;u,m$) parameters as complex numbers. From
\be
[F_{\alpha\beta}]^\dagger &=& [< \frac{1}{2}([a_\alpha ,
a_\beta^\dagger ])^\dagger >_0^\wedge] = [< \frac{1}{2}[a_\beta ,
a_\alpha^\dagger ] >_0^\wedge] = [F_{\beta\alpha}]
\ee
or in matrix form
\be&
\begin{array}{|cc|} -\bar{r}+1/2 & \bar{t} \\ \bar{q} & -\bar{s}+1/2 
\end{array}
=
\begin{array}{|cc|} -r+1/2 & t \\ q & -s+1/2 \end{array}
&\ee
we conclude, that $(q,r,s,t)$ have to be real numbers. From 
\be
u^\dagger &=& < g_{15}^\dagger >_0^\wedge = < (a_1 a_2)^\dagger
>_0^\wedge = - < a_1^\dagger a_2^\dagger >_0^\wedge \nn
&=&  -<g_{16}>_0^\wedge =m
\ee
we obtain that $\Delta_{0}:= um = uu^\dagger \ge 0$ and
analogously from $q^\dagger \simeq g_4^\dagger=g_6\simeq 
t$, $q=q^\dagger=t=t^\dagger$ the relation $\Delta_1 := qt =
qq^\dagger =q^2 \ge 0$. The requirement of hermitecity has thus
furthermore restricted the parameters. The $\Delta_i$ are shifts
induced by the normalization of spin-zero and spin-one eigenvectors.
From (\ref{sol}) we remain with the `gap-equation' (the name will 
become clear later)  
\be\label{eq-s}
&
s=\nu \pm \sqrt{\nu^2 -\Delta_1 + \Delta_0 -w} \, .
&\ee 
The propagator can finally be written as
\be\label{my-prop}
&
F_{\alpha\beta}(\nu,w,\Delta_1-\Delta_0,q) =& \nn
&\begin{array}{|cc|}
-\nu \pm \sqrt{\nu^2-\Delta_1+\Delta_0-w} & q \\
q & -\nu \mp \sqrt{\nu^2-\Delta_1+\Delta_0-w} 
\end{array}.&
\ee

Let us look more closely at the vacuum sector obtained in
our $U(2)$ model. We had already seen, that the space of all
positive vacua is an affine simplex --thereby convex--, spanned
by three extremal states: the Fock state $\omega_F=\omega_{11}$,
the dual Fock state $\omega_{F^*}=\omega_{00}$ and an state
$\omega_E=\omega_{1/2\, 0}$. A general state was decomposed as
\be
\omega_{\nu w} &=& w \omega_F +(1-2\nu+w) \omega_{F^*} +
2(\nu-w) \omega_E.
\ee
Since Bogoliubov transformations do mix Fock and dual
Fock states, they generate an edge of the simplex, which might
also be parameterized as
\be
\omega_{BV} &=& \omega_{1-\rho\, 1-\rho}
\,=\, \rho \omega_{F^*} + (1-\rho) \omega_F, \quad
\rho \in ]0,1[.
\ee
The full space of positive vacua is then a convex combination of
$\omega_{BV}$ and $\omega_E$ with $\lambda \in ]0,1[$
\be\label{states3}
\omega_{\nu w}=
\omega^\prime_{\lambda \rho} &=& \lambda \omega_{BV} + (1-\lambda)
\omega_E \nn
&=& \lambda(1-\rho) \omega_F +\lambda \rho
\omega_{F^*} +(1-\lambda) \omega_E.
\ee
If we now fix a $\rho$ and look at the line of vacua
parameterized by $\lambda$, we get a classical bifurcation
diagram for the solutions of (\ref{sol}) most easily seen if the
$\Delta_i$ are zero, see Fig 1. The quasi free states given by 
$\nu^2=w$ do constitute the borderline of different phases. Since
quasi free states are defined to have no higher correlations,
they separate areas which do posses higher correlations of
possibly different signs. In our case, one area has an
attractive force and an ordered phase, while the other one has
repulsive character. In the area between $\omega_{BV}$ and
$\omega_{\nu\, \nu^2-\Delta_1 + \Delta_2}$ one has a typical
gap-equation, coming from the two real roots of eqn. (\ref{eq-s}), 
where between $\omega_{\nu\, \nu^2-\Delta_1 + \Delta_2}$ and the 
edge state $\omega_E$ this type of solution can not occur in 
(\ref{eq-s}) since $s$ has to be real. 

The $C^*$-algebraic ECAR solution of the QFT hierarchy equations
of BCS-theory can be found in \cite{Ker1,Ker2}, which supports our
terms: vacuum, phase transition, ordered phase, gap-equation, etc.
Copper pairs should be identified with spin-0 states of particle
number --grading-- 2. See also the following discussion.

\begin{figure}[H]
\centering
\includegraphics[width=0.9\textwidth]{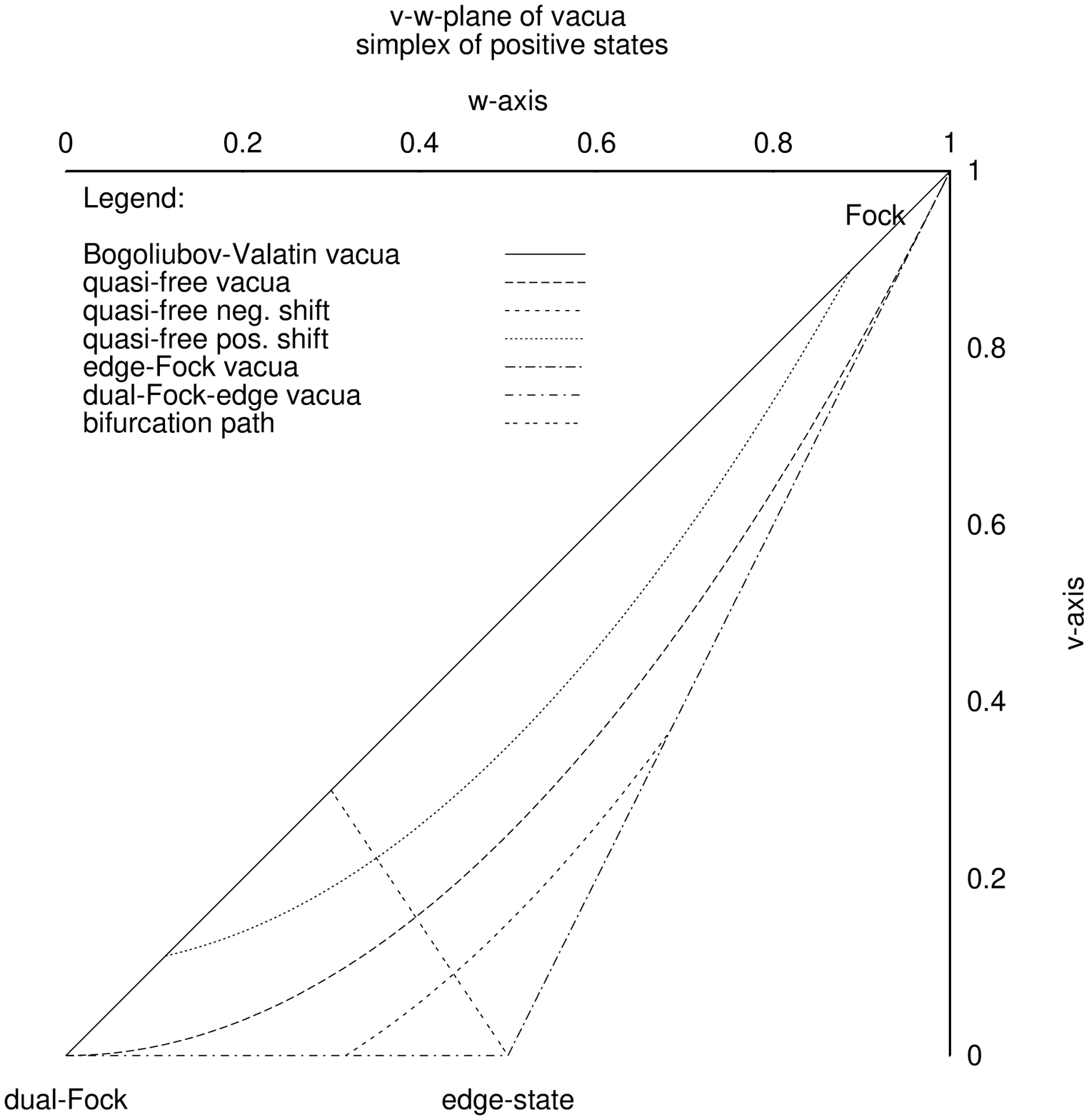}
\caption{%
Simplex of the positive vacua in the $\nu$-$w$-plane.
The interior of the triangle constitutes of all possible non-degenerate
vacua of the $U(2)$-model. The quasi-free states possessing no higher
correlations are described by the parabolas ($\nu$,$\nu^2 \pm \Delta_i$). 
Edge states are pure states and maximally degenerate. Fock, 
dual-Fock and the new edge states span the simplex. States on the 
borderline of the  simplex are also degenerate, especially the 
Bogoliubov-Valatin states which connect the dual-Fock and Fock edges. 
The line connecting the edge-state and a Bogoliubov-Valatin 
state crosses various parabolas of quasi-free states according 
to the relevant shifts. This is the path of the bifurcation diagram
discussed in the text. States below these parabolas do not have a 
condensate, while above it a condensate exists. The shifts $\Delta_i$ 
move the parabolas down and up, thereby increasing or decreasing 
the area of condensate states. In effect, no condensate 
or only condensate states are possible.
}
\label{FIG1}
\end{figure}

\section{\protect\label{SEC-4} Discussion}

\subsection{Relation to Bogoliubov--Valatin transformations}

We want to introduce here the common notation of a Fock vacuum
more explicitly. This will show the equivalence of one of the
above derived vacua to this particular state, but also the restricted
validity of the Fock approach. Especially in QFT, where Haag's 
theorem rules out this possibility.

Let $\mid 0>_F$ ($\mid 0 >_{F^*}$) be a (dual) Fock vector state
which is subjected to the conditions
\be
\prod_{i\in I} a_i \mid 0 >_F &=& 0 \nn
(\, \prod_{i\in I} a_i^\dagger \mid 0 >_{F^*} &=& 0 \,)
\ee
where $I$ contains non-empty ordered subsets of cardinality 
less than or equal to $N$.

The question is, whether this definition is equivalent to our
above description of Fock and dual Fock states. This can be
seen as follows: We calculate, introducing the shorthand
${}_F< 0\mid X \mid 0 >_F \,=\, <X>_F$, the expectation value
\be
0 &=& < a_i^\dagger a_i >_F \, =\, 
<\openone - a_i a_i^\dagger >_F \, =\,
1- \left\{
\begin{array}{l} 
r\quad i=1 \\
s\quad i=2\, .
\end{array} \right.
\ee
Now, $r=s=\nu$ according to (\ref{eq-v1},\ref{g23})
and we end up with
\be
\nu&=& 1.
\ee
From
\be
0 &=& < a_2^\dagger a_1^\dagger a_1 a_2>_F \,=\,
      <a_2^\dagger a_2 - \openone + a_1^\dagger a_1 + 
       a_1 a_2 a_2^\dagger a_1^\dagger >_F \nn
  &=& 0-1+0+w,
\ee
we get
\be
w &=& 1.
\ee
We conclude therefore, that the Fock vacuum is given in our
framework as
\be
\omega_F &=& \omega_{\nu w}\Big\vert_{\nu=1\, w=1} \,=\, \omega_{11},
\ee
whereas the dual-Fock state can be deduced to be given along the
same lines as
\be
\omega_{F^*} &=& 
   \omega_{\nu w}\Big\vert_{\nu=0\, w=0} \,=\, \omega_{00}.
\ee
The state space of the (dual) Fock vacuum is spanned by
the following vectors
\be
{\cal V}_F &=& \{ \openone \mid 0 >_F,
                  a_1^\dagger \mid 0 >_F,
                  a_2^\dagger \mid 0 >_F,
                a_1^\dagger a_2^\dagger \mid 0 >_F \} \nn
{\cal V}_{F^*} &=& \{ \openone \mid 0 >_{F^*},
                  a_1 \mid 0 >_{F^*},
                  a_2 \mid 0 >_{F^*},
                a_1 a_2 \mid 0 >_{F^*} \}.
\ee
Both spaces are four dimensional, but yield different matrix
representations for the group generators e.g. $N$. This shows
explicitly their inequivalence.

In the case of the new edge state, we turn around the direction
of inference to deduce from
\be
\omega_E &=& \omega_{\nu w}\Big\vert_{\nu=1/2\, w=0} 
\,=\, \omega_{1/2\, 0}
\ee
the conditions
\be
a_1 a_2 \mid 0 >_E &=\, 0 \, =&
a_2^\dagger a_1^\dagger \mid 0 >_E.
\ee
The state space of this edge is generated by
\be
{\cal V}_E &=& \{
\openone, a_1, a_2, a_1^\dagger, a_2^\dagger,
\frac{1}{2}(a_1 a_1^\dagger - a_2 a_2^\dagger ),
a_1 a_2^\dagger, a_2 a_1^\dagger \} \mid 0 >_E.
\ee

Note, that neither $a_i$ nor $a_i^\dagger$ annihilate this 
vacuum state. Hence, a linear Bogoliubov-Valatin like 
transformation $c_i = f(a_i,a_j^\dagger)=\alpha_i a_i
+\beta_j a_j^\dagger$ cannot reach this
type of vacuum state. Bogoliubov-Valatin vacua remain on a single
line in our vacuum plane given by $\omega_{BV} = 
\omega_{1-\rho\, 1-\rho}$, see Fig 1. To relate our results 
to the literature, we investigate this connection a little more.

In the theory of superconductivity, Bogoliubov \cite{Bogo},
Valatin \cite{Vala} as Holstein and Primakoff \cite{Hols}
introduced new quasi particles $c,c^\dagger$ via the definition
\be\label{qp}
c_1 &:=& \sqrt{1-h}\, a_1 - \sqrt{h}\, a_2^\dagger \nn
c_2 &:=& \sqrt{1-h}\, a_2 + \sqrt{h}\, a_1^\dagger.
\ee
The transformation is canonical since it fulfils
\be
\{ c_i,c_k^\dagger \} = \delta_{ik}\openone, &\,\,&
\{ c_i,c_k \} \,=\, 0 \,=\, \{ c_i^\dagger, c_k^\dagger \}.
\ee
Furthermore, a new quasi-particle Fock vacuum 
$\mid 0>_{BV} = [\sqrt{1-h} +\sqrt{h} a_1^\dagger a_2^\dagger ] 
 \mid 0>_F$ is introduced, demanding
\be
c_i \mid 0 >_{BV} &=& 0 \quad \forall i.
\ee
Obviously, $\mid 0>_{BV}$ and $\mid 0>_F$ are quite different vector
states of the underlying state space. Furthermore, none of these
states are vacuum states w.r.t. the alternate choice of creation
and annihilation operators. One finds
\be
c_1 \mid 0>_F &=& \sqrt{1-h}\, a_1 + \sqrt{h}\, a_2^\dagger
\mid 0>_F \, =\, \sqrt{h}\mid 0,1 >_F \,\not=\, 0 \nn
a_1 \mid 0>_{BV} &=& a_1 [\sqrt{1-h} + \sqrt{h}\,
a_1^\dagger a_2^\dagger ] \mid 0>_F \,=\, \sqrt{h} \mid 0,1 >_F 
\,\not =\, 0
\ee
etc. in an obvious notation. From this consideration, it is quite
clear, that the notation of a Fock vacuum, i.e. annihilator on a
groundstate equals zero, is highly dependent on the chosen set of
creation and annihilation operators. One cannot speak in this 
terms of a {\it a vacuum}\/ but only of a vacuum w.r.t. $a$,
$a^\dagger$ or $c$, $c^\dagger$ etc. Moreover, the naive mixture of 
creation and annihilation operators spoils particle number conservation 
w.r.t. the old variables, as also the state $\mid 0>_{BV}$ has no definite
$a$-particle number. In consequence, this means that electrons and 
Cooper pairs cannot be sharply measured simultaneously in such a 
framework. Moreover, even if we are able to mimic the 
Bogoliubov-Valatin results, strictly speaking this state is not 
a vacuum in our sense.

An algebraic treatment of states as given above does mix states
and representations in favour of operators and is thus able to
handle mixed states without the described problems, see
\cite{Ker1,Ker2}. However, much more important is the fact that
Bogoliubov--Valatin transformations can only create correlations
due to a mixture of Fock and dual Fock states and fail to be able
to give the full possibilities of the three extremal, or say, pure 
cases. No $\lambda$ mixture can appear as in (\ref{states3}). The
algebraic setting based on GNS techniques in the positive case and
extended to indefinite cases by Clifford algebraic methods is 
thus more general than Bogoliubov-Valatin transformations.

\subsection{BCS propagator according to Fetter Walecka}

To be able to compare our results to the ones usually given in
literature, we derive some further identities. We extend our
theory in the sense that we add a new mode index $k$ to our fields,
as we introduce an independent copy of the algebra for every mode.
This is not harmful, but makes all of our real parameters
functions of $k$. The BCS groundstate is given as \cite{Anderson}
\be
\mid 0 >_{BCS} &=&
\Pi_k [ \sqrt{1-h_k} + \sqrt{h_k}\, a_{k1}^\dagger a_{-k2}^\dagger ]
\mid 0 >_F.
\ee
The connection to the Bogoliubov-Valatin transformation is then
given by
\be\label{qp1}
c_{k1} &:=& u_k\, a_{k1} - v_k\, a_{-k2}^\dagger \nn
c_{k2} &:=& u_k\, a_{k2} + v_k\, a_{-k1}^\dagger,
\ee
compare (\ref{qp}), with the new vacuum defined as Fock vacuum
of the quasi particles. Comparing (\ref{qp1}) and (\ref{qp}) we
get
\be
u_k &=& \sqrt{1-h_k} \nn
v_k &=& \sqrt{h_k}.
\ee
Comparing these quantities with our data, i.e. the vacuum parameters,
we obtain the $u_k$ and $v_k$ in terms of these parameters. From the 
Bogoliubov-Valatin vacuum $\omega_{1-\rho_k\,1-\rho_k}$ we have
\be
c_{k1} &=& \sqrt{\frac{\rho_k-1}{2\rho_k-1}}\, e_{k1}
        - \sqrt{\frac{\rho_k}{2\rho_k-1}}\, e_{-k3} \nn
c_{k2} &=& \sqrt{\frac{\rho_k-1}{2\rho_k-1}}\, e_{k2}
        + \sqrt{\frac{\rho_k}{2\rho_k-1}}\, e_{-k4},
\ee
and thus
\be
\label{eq91}
\nu_k \,=\, w_k &=& 1-\rho_k \nn
u_k &=& \sqrt{\frac{\rho_k-1}{2\rho_k-1}} \nn
v_k &=& \sqrt{\frac{\rho_k}{2\rho_k-1}},
\ee
iff we set the shifts $u=m=0$. From $\nu_k=w_k=1-\rho_k$ we can
derive all of our propagator data etc. in terms of $\rho_k$
and finally in terms of the $u_k,v_k$ parameters of the
Bogoliubov-Valatin transformation. We can thus identify
the results given in literature with those given by Clifford
algebraic considerations. 

We follow Fetter and Walecka (FW) \cite{FetterWalecka}. The model
Hamiltonian is given in eq. (51.1), which can be written formally
as
\be
H &=& \epsilon g_2 + g g_4
\ee
in our notation. This equation is solved using the mean field
ansatz. It is convenient to introduce the following functions
(FW eq.s 51.10, 51.12a, 51.12b, 51.14)
\be
{\cal G}(x,x^\prime) &:=& 
   - < T[\psi_{k\uparrow} \psi^\dagger_{k\uparrow}]> 
\quad (\cong\, <a_i a_i^\dagger >_{\nu w} \,)\nn
{\cal F}(x,x^\prime) &:=& 
   - < T[\psi_{k\uparrow} \psi_{k\downarrow}]>
\quad (\cong\, <a_1 a_2 >_{\nu w} \,)\nn
{\cal F}^\dagger(x,x^\prime) &:=& 
   - < T[\psi^\dagger_{k\downarrow} \psi^\dagger_{k\uparrow}]>
\quad (\cong\, <a_2^\dagger a_1^\dagger >_{\nu w} \,)\nn
\Delta(x) &:=& g {\cal F}(x,x).
\ee
$\Delta$ is the gap-function, which afterwards describes the lowering 
of energy in the superconducting phase. In our above notation, we had
$F$ as propagator, which corresponds to the Green function ${\cal G}$
while ${\cal F}$ and ${\cal F}^\dagger$ have to be identified with 
the shift $\Delta_0=uu^\dagger$. This renders $\vert u \vert$ to be
the gap for a single mode. However, as we saw in the Clifford 
geometric approach, this parameter is not responsible for the 
general structure of the vacua in the $\nu$-$w$-plane, but induces
only a shift. Unfortunately the Bogoliubov-Valatin transformation 
and the BCS theory loose this connection by mixing operators and 
not properly mixing the states. If we introduce 
$\xi_k = \epsilon_k - \mu = \hbar^2k^2/2m -\mu$ 
(FW eq. 37.24 and 51.29) as the energy relative
to the chemical potential, one can give the above functions in
momentum space as (FW eq.s 51.30a, 51.30b)
\be
{\cal G}(k,\omega) &=&
\frac{-\hbar(i\hbar \omega + \xi_k)}{
 \hbar^2 \omega^2 + \xi_k^2 +\mid \Delta \mid^2} \nn
{\cal F}(k,\omega) &=&
\frac{\hbar \Delta^*}{
 \hbar^2 \omega^2 + \xi_k^2 +\mid \Delta \mid^2}
\ee
where the gap $\Delta$ is real in absence of an external field.
As a last step, one has to specify these functions as functions
of the parameters of the Bogoliubov-Valatin transformation. 
This yields (FW eq.s 51.31--51.34)
\be
E_k &=& \sqrt{\xi_k^2 + \Delta^2} \nn
v_k u_k &=& \frac{\Delta}{2E_k} \nn
v_k^2 &=& 1- u_k^2 \,=\, \frac{1}{2}(1- \frac{\xi_k}{E_k}) \nn
{\cal G}(k,\omega) &=&
\frac{u_k^2}{i\omega_k - E_k/\hbar} +
\frac{v_k^2}{i\omega_k + E_k/\hbar} \nn
{\cal F}(k,\omega) &=& {\cal F}^\dagger(k,\omega) \nn
&=&
u_k v_k\left[
\frac{1}{i\omega_k - E_k/\hbar} -\frac{1}{i\omega_k + E_k/\hbar}
\right].
\ee
Inserting our expressions for the $u_k,v_k$ in terms of the 
$\rho_k$ (\ref{eq91})we are able to get all of the results as usual.
However, this discussion makes it quite clear, that the common
treatment of BCS theory involves several peculiar steps. First
of all, it was unveiled that the BCS and/or Bogoliubov-Valatin
treatment is not able to include the third edge state and deals
thus with a restricted theory. Secondly, the BCS and 
Bogoliubov-Valatin treatment spoils particle number conservation
due to the naive concept of vacua involved. A more elaborate 
treatment which mixes the algebraic states and not the operators
can surmount this difficulty. Thirdly, the gap is calculated
afterwards by a consistency condition and/or a variational method.
The algebraic theory however fixes these parameters uniquely
from propagator data and thus ultimatively from the chosen 
Hamiltonian. This can easily include the thermodynamical
behaviour of the theory, if the model Hamiltonian is explicitly
known in terms of temperature dependent couplings etc.
One ends up with a theory such as
\be
H &=& \alpha(T,B,\cdots)\openone +
      \epsilon(T,B,\cdots) g_2 + 
       g(T,B,\cdots) g_4.
\ee
Playing with the external parameters $T,B,\cdots$ we can move 
through the vacuum plane, since $\nu$ and $w$ become functions
of $T,B,\cdots$. This explains also, in which way a phase transition
can be experimentally obtained. However, the dependence of the coupling
constants on these thermodynamical variables can be derived only
from a more subtle microscopic theory which is beyond BCS theory.
One should also compare the treatment of superconductivity in 
\cite{StuBo}.

\subsection{Summary}

By embedding CAR algebras and Extended CAR algebras into a
generalized Clifford geometric framework it was possible to
unveil several new aspects. This fact is based on an intrinsic
algebraization of states in Clifford algebras of multivectors.
If we have a duality pairing $<.,.>$ and an exterior product
$\wedge$, we define by
\be
 <a \con b , c> &=& <b, a\mytil \wedge c>
\ee
the contraction as the dual product of the wedge w.r.t. the
given duality pairing. On the other hand, we might fix a
contraction and a wedge, thereby defining a duality pairing.
This cannot be achieved if the positivity requirement is
considered, since the positive states constitute only a convex
affine set, which is i.g. not linear. Furthermore, we can not
find a representation of the full algebraic setting in a
positive nondegenerate space of appropriate dimension, which is
seen from the GNS construction and Clifford algebra theory.
However, we saw that ECAR methods are able to describe correctly
the vacuum sector, but looses thereby the {\it constructive}\/
relation between the propagator matrix elements and the vacuum
parameterization. This relation is uncovered by introducing
Clifford algebras of multivectors or quantum Clifford algebras, 
which possess possibly an antisymmetric part in the contraction. 
In general, this introduces $n(n-1)/2$ parameters, which can be 
used to describe inequivalent vacua. However, since one has already 
chosen a basis, these propagator matrix elements are basis dependent
and thereby observer dependent, so that not all of them describe
properties of vacua. We succeeded in our $U(2)$-model, to give
the explicit functional relations of the vacuum parameters
$\nu,w$ in terms of the contraction data. As we were able to
parameterize all possible propagator we can describe all 
possible dynamics by five parameters:  
$\nu, w, \Delta_0, \Delta_1; q$.

We were able to give a complete classification of the vacuum
sector of positive states, including a phase transition. We
obtained, furthermore, shifts in the gap-equation induced by
spin-one and spin-zero eigenvectors. No additional ordering parameter
or potential was needed to obtain this transition,
which occurs for {\it purely algebraic reasons}.\/ 
This ordering parameter is replaced by the Higgs field in elementary 
particle theory. Moreover, we
did not even need to define a specific dynamics but used the 
propagator as a parameter. This renders the vacuum structure to be
universal to any models or choices of Hamiltonians possible 
within our framework. 

Since our method can formally be generalized to QFT in a
straightforward manner \cite{Fau-pos,Fau-diss} and even to
symplectic Clifford algebras, which are related to CCR algebras
and bosons \cite{Fau-pos}, we conclude that symmetry breaking
might occur without a Higgs field. The Higgs field of elementary
particle physics would therefore {\it ad hoc}\/ reintroduce the lost 
connection between contraction data and the vacua.

Our model shows very clearly that phase transitions do not 
depend on an infinite number of particles or a thermodynamic 
limit.

The relation between operators and observables has been
clarified by explicit calculations, which lead e.g. to an {\it
additive}\/ renormalization of eigenmonomials in the generators.
Furthermore, our method is {\it not}\/ restricted to the
positive case. Positivity was only considered for comparing our
results with conventional calculations. If we allow indefinite
states, we have to move to the geometrical point of view, since
the statistical interpretation breaks down. In QCD and
non-linear spinor field theories ghost fields occur, e.g. the
Faddeev-Popov-fields, which are needed during interaction, but
not in the initial and final states, which are subjected to a
statistical interpretation. We hope that the geometric concepts,
valid in indefinite situations, will be useful in the investigation
of this situation.

Further developments of the theory shall include explicitly
dynamical models with special Hamiltonians. This will then allow
us to calculate the propagator matrix elements and therewith
the correct and consistent ground state. This will be 
interesting in non-linear dynamics which intrinsically
specifies all data.

\subsection{Epilogue}

Our method as presented in this article remains in the area of algebraic
developments. However, recent research has unveiled a deeper link of this
method to Hopf algebras. Starting from a pair, product and co-product on
a space, one can define a convolution product which might be a bigebra or
Hopf gebra if additional requirements as a crossing or an antipode are
imposed. One finds, that a product on the dual space induces a co-product
on the original space. Since we have induced such a duality via our
unique vacuum state and such a duality is already employed, the question
arises, if we are dealing here silently with a Hopf gebra. One notes,
that only with the help of quantum Clifford algebras it is possible to
stay with a Hopf gebra structure \cite{FauOzi}. If the scalar and
co-scalar product are mutually inverse one of another, then only a
Clifford convolution algebra can be defined. However, for Grassmann
algebras a unique and natural Hopf algebra structure is easily established.
This structure allows one to find deeper arguments what is going on e.g.
in re-ordering in QFT \cite{Fau-wick}. There it was demonstrated that our
vacuum is nothing but the co-unit of the involved Hopf gebra and that a
process called cliffordization \cite{RotSte} connects the different sets
of bases which we have parameterized by $v$ or $v,w$ in our examples. This
outcome supports our point that dynamics should select a unique `vacuum'
state, i.e. a unique co-unit. Having the power of the Hopf gebra at our
disposal, one can turn the usual arguments in QFT upside down and seek
for {\it axioms} which provide models which can be treated as QF
theories. In \cite{Fau-grp00a} we gave a first account how such axioms
might be set up to avoid singularities in QFT by algebraic design. The
present work is thus the foundation of such an algebraic axiomatization
of quantum physics, which renders various QF theories to be instances of
sets of algebraic axioms.

\section{Acknowledgements}

This paper was inspired by the possibility to play with
``Clifford'' Ver. 4, a Maple V Release 5 add-on programmed by
Prof. Rafal Ablamowicz, Tennessee \cite{Abla}. His generous
help in extending the features, collected in the Cli4plus package, 
were a valuable help.
The support of Mrs. Ursula Wieland is gratefully acknowledged. 
My gratitude is expressed to Ms. Kirsten Magee for improving
my English.

\end{document}